# A kind of quantum dialogue protocols without information leakage assisted by auxiliary quantum operation

Ling-Yun Huang, Tian-Yu Ye*

College of Information & Electronic Engineering, Zhejiang Gongshang University, Hangzhou 310018, P.R.China

Corresponding author's e-mail：happyyty@aliyun.com

**Abstract:** In this paper, a kind of quantum dialogue (QD) protocols without information leakage assisted by quantum operation is proposed. The participant in charge of preparation can automatically know the collapsed states after quantum operation performed on the prepared quantum states. The other participant is able to know the collapsed states derived from the prepared quantum states through quantum measurement. The information leakage problem is avoided by means of imposing auxiliary quantum operation on the prepared quantum states.

**Keywords:** Quantum dialogue(QD); information leakage; quantum operation.

## 0 Introduction

As a special kind of quantum secure direct communication (QSDC), quantum dialogue(QD) aims to accomplish the bidirectional transmission of secret messages between two participants. Since it was independently suggested by Zhang et al.[1-2] and Nguyen[3] in 2004, a lot QD protocols[4-13] had been proposed by many scholars. However, Tan and Cai [14] found out in 2008 that there may exist "classical correlation" in QD. In the meanwhile, Gao et al. [15-16] also pointed out in 2008 that "information leakage" may happen in QD. Since then, scholars began to pay special attentions on the information leakage problem in QD. As a result, many good information leakage resistant QD protocols [17-23] were put forward subsequently. In this paper, we also concentrate on the issue of information problem in QD. With the help of quantum operation, we propose a kind of QD protocols without information leakage. In this kind of QD protocols, the participant in charge of preparation can automatically know the collapsed states after quantum operation performed on the prepared quantum states. Moreover, the other participant is able to know the collapsed states derived from the prepared quantum states through quantum measurement. The information leakage problem is avoided by means of imposing auxiliary quantum operation on the prepared quantum states. The remaining part of this paper is organized as follows: the QD protocol using Bell state is illustrated in section 1; the QD protocol using W state is illustrated in section 2; the QD protocol using four-particle GHZ state is illustrated in section 3; section 4 compares our protocols with previous information leakage resistant QD protocols; finally, the conclusion is made in section 5.

## 1 QD protocol using Bell state
### 1.1 Protocol description

Before beginning the protocol description, let us illustrate the effect from the controlled-not (CNOT) operation performed on Bell states. Four Bell states are defined as follows:

$$\left|\phi^+\right\rangle_{ab} = \frac{1}{\sqrt{2}}\left(\left|00\right\rangle + \left|11\right\rangle\right)_{ab}, \tag{1}$$

$$\left|\phi^-\right\rangle_{ab} = \frac{1}{\sqrt{2}}\left(\left|00\right\rangle - \left|11\right\rangle\right)_{ab}, \tag{2}$$

$$\left|\psi^+\right\rangle_{ab} = \frac{1}{\sqrt{2}}\left(\left|01\right\rangle + \left|10\right\rangle\right)_{ab}, \tag{3}$$

$$\left|\psi^-\right\rangle_{ab} = \frac{1}{\sqrt{2}}\left(\left|01\right\rangle - \left|10\right\rangle\right)_{ab}. \tag{4}$$

When the CNOT operation is performed on the four Bell states with particle $a$ as the control qubit and particle $b$ as the target qubit, respectively, it follows:

$$U_{CNOT}^{ab} \otimes \left|\phi^+\right\rangle_{ab} = U_{CNOT}^{ab} \otimes \frac{1}{\sqrt{2}}\left(\left|00\right\rangle + \left|11\right\rangle\right)_{ab} = \frac{1}{\sqrt{2}}\left(\left|00\right\rangle_{ab} + \left|10\right\rangle_{ab}\right) = \left|+\right\rangle_a \left|0\right\rangle_b, \tag{5}$$

$$U_{CNOT}^{ab} \otimes \left|\phi^-\right\rangle_{ab} = U_{CNOT}^{ab} \otimes \frac{1}{\sqrt{2}}\left(\left|00\right\rangle - \left|11\right\rangle\right)_{ab} = \frac{1}{\sqrt{2}}\left(\left|00\right\rangle_{ab} - \left|10\right\rangle_{ab}\right) = \left|-\right\rangle_a \left|0\right\rangle_b, \tag{6}$$

$$U_{CNOT}^{ab} \otimes \left|\psi^+\right\rangle_{ab} = U_{CNOT}^{ab} \otimes \frac{1}{\sqrt{2}}\left(\left|01\right\rangle + \left|10\right\rangle\right)_{ab} = \frac{1}{\sqrt{2}}\left(\left|01\right\rangle_{ab} + \left|11\right\rangle_{ab}\right) = \left|+\right\rangle_a \left|1\right\rangle_b, \tag{7}$$

$$U_{CNOT}^{ab} \otimes \left|\psi^-\right\rangle_{ab} = U_{CNOT}^{ab} \otimes \frac{1}{\sqrt{2}}\left(\left|01\right\rangle - \left|10\right\rangle\right)_{ab} = \frac{1}{\sqrt{2}}\left(\left|01\right\rangle_{ab} - \left|11\right\rangle_{ab}\right) = \left|-\right\rangle_a \left|1\right\rangle_b, \tag{8}$$

where $|\pm\rangle = \frac{1}{\sqrt{2}}(|0\rangle \pm |1\rangle)$. Apparently, after the CNOT operation, each of the four Bell states collapses into two independent single particles without any entanglement.

Suppose that Alice has a secret message consisting of $2N$ bits $\{i_1, i_2, \cdots, i_n, \cdots, i_{2N}\}$ and Bob owns a secret message consisting of $2N$ bits $\{j_1, j_2, \cdots, j_n, \cdots, j_{2N}\}$, where $i_n, j_n \in \{0,1\}, n \in \{1, 2, \cdots, 2N\}$. They agree on in advance that the unitary operation $I$ ($i\sigma_y$) stands for bit 0 (1), where $I = |0\rangle\langle 0| + |1\rangle\langle 1|$ and $i\sigma_y = |0\rangle\langle 1| - |1\rangle\langle 0|$.

The proposed protocol consists of the following steps.

**Step 1: Preparation.** Alice produces $N + \delta_1 + \delta_2$ EPR pairs randomly in one of the above four Bell states, where the $N$ EPR pairs and the $\delta_1 + \delta_2$ EPR pairs are used for message transmission and security check, respectively. Here, the $N + \delta_1 + \delta_2$ EPR pairs are denoted as $\{[P_1(a), P_1(b)], [P_2(a), P_2(b)], \cdots, [P_n(a), P_n(b)], \cdots, [P_{N+\delta_1+\delta_2}(a), P_{N+\delta_1+\delta_2}(b)]\}$, where the subscripts indicate the orders of EPR pairs in sequence, while $a$ and $b$ stand for two particles in each EPR pair, respectively. Alice picks out particle $a$ from each EPR pair to form sequence $A$, i.e., $A = [P_1(a), P_2(a), \cdots, P_n(a), \cdots, P_{N+\delta_1+\delta_2}(a)]$. Likewise, the remaining particles form sequence $B$, i.e., $B = [P_1(b), P_2(b), \cdots, P_n(b), \cdots, P_{N+\delta_1+\delta_2}(b)]$.

**Step 2: The first transmission and the first security check.** Alice sends sequence $A$ to Bob in the manner of block transmission [24]. After Bob announces Alice that he has received sequence $A$, they work together to check eavesdropping in the way same to the first security check of Deng et al.'s two-step QSDC[25]: (1) Alice informs Bob of the positions of $\delta_1$ checking particle $a$ in sequence $A$; (2) Bob randomly selects one of the two measuring bases, Z-basis ($\{|0\rangle, |1\rangle\}$) and X-basis ($\{|+\rangle, |-\rangle\}$), to measure $\delta_1$ checking particle $a$, and tells Alice his measurement bases and measurement results; (3) Alice chooses the same measurement bases as Bob's to measure the corresponding $\delta_1$ checking particle $b$; (4) Through comparing her own measurement results with Bob's, Alice can ascertain whether the quantum channel is secure or not. If there is no eavesdropping, their measurement results should have deterministic correlations so that they carry on the next step. Otherwise, they abort the communication.

**Step 3: The second transmission and the second security check.** Alice (Bob) discards the $\delta_1$ checking particle $b$ ($a$) in sequence $B$ ($A$). Then Alice sends sequence $B$ to Bob in the manner of block transmission [24]. After Bob announces Alice that he has received sequence $B$, they collaborate to implement the second security check in the way same to the second security check of Deng et al.'s two-step QSDC[25]: (1)Alice tells Bob the positions of the $\delta_2$ checking EPR pairs in sequences $A$ and $B$; (2) Bob performs the Bell-basis measurement on each of the $\delta_2$ checking EPR pairs and announces his measurement result to Alice; (3)Alice judges whether the quantum channel is secure or not by comparing the initial state of each of the $\delta_2$ checking EPR pairs with Bob's measurement result. If they are identical, the quantum channel turns out to be safe so that they go on the next step. Otherwise, they have to stop the communication.

**Step 4: Bob's encoding.** After Bob discards the $\delta_2$ checking EPR pairs, he performs a CNOT operation on each of the remaining $N$ EPR pairs with particle $a$ as the control qubit and particle $b$ as the target qubit. According to Eqs.(5-8), Alice can automatically know the states of each particle $a$ and each particle $b$ after the CNOT operation, since she prepares each of the $N$ EPR pairs by herself. As to Bob, in order to know the state of each particle $a$ ($b$) after the CNOT operation, he makes the X-basis (Z-basis) measurement on it. According to his X-basis (Z-basis) measurement result, Bob reproduces a new particle $a$ ($b$) on which state measurement is not performed. Then Bob mixes all new particles $a$ and $b$ together to form sequence $C$, i.e., $C = [P_1(c), P_2(c), \cdots, P_n(c), \cdots, P_{2N}(c)] (c \in \{a, b\})$, and keeps a record of their exact positions in sequence $C$. Afterward, Bob encodes his own one-bit secret on each particle $c$ by performing one of the two unitary operations $\{I, i\sigma_y\}$. Thus, sequence $C$ is changed into sequence $C'$, i.e., $C' = \{U_1^\beta P_1(c), U_2^\beta P_2(c), \cdots, U_n^\beta P_n(c), \cdots, U_{2N}^\beta P_{2N}(c)\}$, where $U_1^\beta, U_2^\beta, \cdots, U_{2N}^\beta \in \{I, i\sigma_y\}$.

**Step 5: The third transmission and the third security check.** For the third security check, Bob adopts the decoy photon technology[26-27]. That is, Bob produces $\delta_3$ decoy photons randomly in one of the four states $\{|0\rangle, |1\rangle, |+\rangle, |-\rangle\}$, and inserts them randomly into sequence $C'$. Then, Bob sends sequence $C'$ to Alice in the manner of block transmission [24]. After Alice tells Bob that she has received sequence $C'$, they work together to implement the third security check as follows: (1) Bob informs Alice of the positions of the $\delta_3$ decoy photons; (2)Bob tells Alice the right measurement base for measuring each of the $\delta_3$ decoy photons; (3) Alice uses the measurement base Bob told to measure each of the $\delta_3$ decoy photons, and tells her measurement result to Bob; (4) Through comparing the initial state of each of the $\delta_3$ decoy photons with Alice's measurement result, Bob can ascertain whether the channel is secure or not. If there is no eavesdropping, they will carry on the next step; otherwise, they stop the communication.

**Step 6: Alice's encoding and bidirectional communication.** After Alice discards the $\delta_3$ decoy photons, she encodes her own one-bit secret on each particle $c$ in sequence $C'$ by performing one of the two unitary operations $\{I, i\sigma_y\}$. As a result, sequence $C'$ is changed into sequence $C''$, i.e., $C'' = \{U_1^\alpha U_1^\beta P_1(c), U_2^\alpha U_2^\beta P_2(c), \cdots, U_n^\alpha U_n^\beta P_n(c), \cdots, U_{2N}^\alpha U_{2N}^\beta P_{2N}(c)\}$, where $U_1^\alpha, U_2^\alpha, \cdots, U_{2N}^\alpha \in \{I, i\sigma_y\}$. Bob tells Alice the exact positions of particles $a$ and $b$ in sequence $C$ he recorded. Consequently, Alice can select the right measurement base to measure each particle $c$ in sequence $C''$, since both of the unitary operations $I$ and $i\sigma_y$ do not alter the bases of particles $a$ and $b$. Then, Alice announces all of her measurement results to Bob. Accordingly, with his own unitary operations and his knowledge about the states of particle $a$ ($b$) after the CNOT operation from his own X-basis (Z-basis) measurements, Bob can extract Alice's secret messages easily. Likewise, Alice can also easily read out Bob's secret messages through her own unitary operations and her knowledge about the states of particle $a$ ($b$) after the CNOT operation from her preparation of EPR pairs.

Here, an example is given to further explain the above protocol. Assume that the first Bell state Alice produces is in the state of $|\phi^+\rangle_{a_1 b_1}$. Suppose that by using the first Bell state $|\phi^+\rangle_{a_1 b_1}$, Alice and Bob want to transmit bits '**10**' and '**01**' to each other, respectively. After having particles $a_1$ and $b_1$ in hand, Bob imposes a CNOT operation on $|\phi^+\rangle_{a_1 b_1}$ with particle $a_1$ as the control qubit and particle $b_1$ as the target qubit. Consequently, the $|\phi^+\rangle_{a_1 b_1}$ evolves as follows:

$$U_{CNOT}^{a_1 b_1} \otimes |\phi^+\rangle_{a_1 b_1} = |+\rangle_{a_1} |0\rangle_{b_1}. \tag{9}$$

Then Bob makes the X-basis (Z-basis) measurement on particle $a_1$ ($b_1$) to know its state. According to his X-basis (Z-basis) measurement result, Bob reproduces a new particle $a_1$ ($b_1$) on which state measurement is not performed. Without loss of generality, assume that particle $b_1$ is located ahead of particle $a_1$ in sequence $C$. Then, particles $b_1$ and $a_1$ evolves as follows:

$$\begin{cases} |0\rangle_{b_1} \Rightarrow I^\beta \otimes |0\rangle_{b_1} = |0\rangle_{b_1} \Rightarrow i\sigma_y^\alpha \otimes |0\rangle_{b_1} = |1\rangle_{b_1} \\ |+\rangle_{a_1} \Rightarrow i\sigma_y^\beta \otimes |+\rangle_{a_1} = |-\rangle_{a_1} \Rightarrow I^\alpha \otimes |-\rangle_{a_1} = |-\rangle_{a_1} \end{cases}. \tag{10}$$

Bob tells Alice the exact positions of particles $b_1$ and $a_1$ in sequence $C$ he recorded. Accordingly, Alice selects Z-basis (X-basis) to measure particle $b_1$ ($a_1$), and announces her measurement results to Bob. With Alice's announced measurement result of particle $b_1$ ($a_1$), his own unitary operation $I^\beta$ ($i\sigma_y^\beta$) and his knowledge about the initial state of particle $b_1$ ($a_1$) after the CNOT operation, Bob can read out that Alice's first (second) bit is **1**(**0**). Likewise, according to her own unitary operation $i\sigma_y^\alpha$ ($I^\alpha$) and her knowledge about the initial state of particle $b_1$ ($a_1$) after the CNOT operation, Alice can also read out that Bob's first (second) bit is **0**(**1**).

## 1.2 Security analysis

We also take the above case as an example to analyze the information leakage problem. Bob performs a CNOT operation on $|\phi^+\rangle_{a_1 b_1}$ with particle $a_1$ as the control qubit and particle $b_1$ as the target qubit. Since she prepares the $|\phi^+\rangle_{a_1 b_1}$ by herself, Alice can automatically know the states of particles $a_1$ and $b_1$ after the CNOT operation from Eq.(9). Moreover, Bob knows the state of particle $a_1$ ($b_1$) after the CNOT operation by making the X-basis (Z-basis) measurement on it. Therefore, it is unnecessary for both Alice and Bob to publicly announce the states of particles $a_1$ and $b_1$ after the CNOT operation. As a result, Eve has no chance to know the states of particles $a_1$ and $b_1$ after the CNOT operation. Without loss of generality, take particle $a_1$ for example. After hearing from Alice's announced measurement result $|-\rangle_{a_1}$, if Eve guesses the initial state of particle $a_1$ after the CNOT operation is $|+\rangle_{a_1}$ ($|-\rangle_{a_1}$), Alice and Bob's second bits will be **01** or **10** (**00** or **11**). As a result, the quantum channel includes $-\sum_{i=1}^{4} p_i \log_2 p_i = -4 \times \frac{1}{4} \log_2 \frac{1}{4} = 2$ bit information for Eve. Therefore, no information is leaked out to Eve.

Then, we turn to analyze the security towards Eve's active attacks. In the second transmission, Alice sends sequence $B$ to Bob. In fact, during this transmission, Eve cannot distinguish a Bell state by only intercepting particle $b$, since it is in a complete mixed state. As a result, Eve can only disturb this transmission and obtain nothing useful. Therefore, the security about the transmission of the prepared EPR pairs depends on the first transmission. The first security check uses the entanglement correlation between two particles from a Bell state to detect an eavesdropper, same to the first security check of Deng et al.'s two-step QSDC[25]. Also, the effectiveness of this method towards Eve's active attacks, such as the intercept-resend attack, the measure-resend attack and the entangle-measure attack, has been fully discussed in Refs.[13,17,19,20]. In the third transmission, Bob sends sequence $C'$ to Alice. The third security check adopts the decoy photon technology [26-27], i.e., uses decoy photons randomly in one of the four states $\{|0\rangle, |1\rangle, |+\rangle, |-\rangle\}$ to detect an eavesdropper. Also, the effectiveness of this method towards Eve's above active attacks has been fully analyzed in Refs.[10,13].

### 1.3 Calculation of information-theoretical efficiency

The information-theoretical efficiency of quantum communication protocol defined by Cabello[28] is $\eta = \frac{b_s}{q_t - b_t}$, where $b_s$, $q_t$ and $b_t$ are the expected secret bits received, the qubits used and the classical bits exchanged between Alice and Bob, respectively. Without loss of generality, take the above case as example too. The $|\phi^+\rangle_{a_1 b_1}$ can be used for transmitting Alice's two bits and Bob's two bits with two bits needed for announcing Alice's measurement results on the final states of particles $a_1$ and $b_1$. Thus it follows that $b_s = 4$, $q_t = 2$ and $b_t = 2$, making $\eta = \frac{4}{2+2} \times 100\% = 100\%$.

## 2 QD protocol using W state
### 2.1 Protocol description

Before beginning the protocol description, let us illustrate the effect from the exchange transformation performed on W state. The W state this protocol involves is

$$|W\rangle_{abc} = \frac{1}{2}(|001\rangle + |010\rangle - |100\rangle + |111\rangle)_{abc} = \frac{1}{\sqrt{2}}(|0\rangle_a |\psi^+\rangle_{bc} - |1\rangle_a |\phi^-\rangle_{bc}). \tag{11}$$

Apparently, if the Z-basis measurement and the Bell-basis measurements are imposed on particle $a$ and particles $b$ and $c$, respectively, the above W state will collapse into $|0\rangle_a |\psi^+\rangle_{bc}$ and $|1\rangle_a |\phi^-\rangle_{bc}$ each with probability 50%. The exchange transformation used in this protocol is defined as

$$U_{EX} = |00\rangle\langle 00| + |01\rangle\langle 10| + |10\rangle\langle 01| + |11\rangle\langle 11|. \tag{12}$$

After the exchange transformation is performed on particles $a$ and $b$, we obtains

$$U_{EX}^{ab} \otimes |W\rangle_{abc} = \frac{1}{2}(|001\rangle + |100\rangle - |010\rangle + |111\rangle)_{abc} = \frac{1}{\sqrt{2}}(|0\rangle_a |\psi^-\rangle_{bc} + |1\rangle_a |\phi^+\rangle_{bc}). \tag{13}$$

Obviously, after the exchange transformation, the W state collapses into $|0\rangle_a |\psi^-\rangle_{bc}$ and $|1\rangle_a |\phi^+\rangle_{bc}$ each with probability 50%.

Suppose that both Alice and Bob have a two-bit for their own. They reach a consensus in advance that : (1) At Bob's site, the four Bell states $|\psi^-\rangle$, $|\phi^+\rangle$, $|\psi^+\rangle$ and $|\phi^-\rangle$ stand for the classical bits 00, 01, 10 and 11, respectively; (2) At Alice's site, the four unitary operations $U_{00}$, $U_{01}$, $U_{10}$ and $U_{11}$ represent the classical bits 00, 01, 10 and 11, respectively, where $U_{00} = I = |0\rangle\langle 0| + |1\rangle\langle 1|$, $U_{01} = \sigma_x = |0\rangle\langle 1| + |1\rangle\langle 0|$, $U_{10} = \sigma_z = |0\rangle\langle 0| - |1\rangle\langle 1|$ and $U_{11} = i\sigma_y = |0\rangle\langle 1| - |1\rangle\langle 0|$.

The proposed protocol is described in detail as follows. As we pay our major attention on the information leakage problem of QD, we omit the description of security checks here, which can be regarded to be independent from the communication process to a great extent. Alice produces one W state in the state of $|W\rangle_{abc}$, and sends particles $a$ and $b$ to Bob. Bob encodes his own two-bit in the following way: (a) If his two-bit is **00** or **01**, he performs $U_{EX}$ on particles $a$ and $b$ first. Then, he uses Z-basis to measure particle $a$. As a result, he can know the state particles $b$ and $c$ are collapsed into. If particles $b$ and $c$ are in the state of $|\phi^+\rangle_{bc}$, he will perform $U_{11}$ on particle $b$ for transmitting **00** to Alice (i.e., $|\phi^+\rangle_{bc}$ is converted into $|\psi^-\rangle_{bc}$), and do nothing on particle $b$ for transmitting **01** to Alice (i.e., $|\phi^+\rangle_{bc}$ is kept unchanged). On the other hand, if particles $b$ and $c$ are in the state of $|\psi^-\rangle_{bc}$, the opposite actions can be taken by him for transmitting his own two-bit to Alice; (b) If his two-bit is **10** or **11**, he does not perform any $U_{EX}$ on particles $a$ and $b$. He uses Z-basis to measure particle $a$. As a result, he can know the state particles $b$ and $c$ are collapsed into. If particles $b$ and $c$ are in the state of $|\phi^-\rangle_{bc}$, he will perform $U_{11}$ on particle $b$ for transmitting **10** to Alice (i.e., $|\phi^-\rangle_{bc}$ is converted into $|\psi^+\rangle_{bc}$), and do nothing on particle $b$ for transmitting **11** to Alice (i.e., $|\phi^-\rangle_{bc}$ is kept unchanged). On the other hand, if particles $b$ and $c$ are in the state of $|\psi^+\rangle_{bc}$, the opposite actions can be taken by him for transmitting his own two-bit to Alice. After finishing his encoding, Bob sends particle $b$ to Alice. After obtaining particle $b$, Alice encodes her own two-bit by performing one of the four unitary operations $\{U_{00}, U_{01}, U_{10}, U_{11}\}$ on particle $b$. Then, Alice uses Bell-basis to measure particles $b$ and $c$, and announces her measurement result to Bob publicly. Accordingly, Bob can read out Alice's two-bit by virtue of Alice's announcement and his own operations. Similarly, Alice can easily infer Bob's two-bit from her own unitary operation.

Particularly, a concrete example is given to further explain the above protocol. Suppose that Alice and Bob expect to transmit bits **10** and **01** to each other, respectively. After receiving particles $a$ and $b$, Bob imposes $U_{EX}$ on particles $a$ and $b$. Consequently, the $|W\rangle_{abc}$ evolves in a manner shown as Eq.(13). Then Bob performs Z-basis measurement on particle $a$ to know its state. Suppose that after Bob's measurement, particle $a$ is collapsed into the state of $|1\rangle_a$. Bob can easily recognize that particles $b$ and $c$ are collapsed into the state of $|\phi^+\rangle_{bc}$. As the $|\phi^+\rangle_{bc}$ can stand for Bob's two-bit **01** at his site, it is unnecessary for him to perform any unitary operation on particle $b$. Hence, Bob sends particle $b$ to Alice directly. After obtaining particle $b$, Alice imposes $U_{10}$ on it. As a result, particles $b$ and $c$ evolves into

$$U_{10}^b \otimes |\phi^+\rangle_{bc} = |\phi^-\rangle_{bc}. \tag{14}$$

Then Alice uses Bell-basis to measure particles $b$ and $c$, and announces her measurement result to Bob. With Alice's announced measurement result, Bob can read out that Alice's two-bit is **10**. Analogously, according to her own unitary operation $U_{10}$, Alice can know that Bob's two-bit is **01**.

## 2.2 Security analysis

The information leakage problem of the above protocol is analyzed here. Apparently, Eve does not know the state of particles $b$ and $c$ exactly at both Bob's and Alice's site. At Bob's site, there are four possibilities for the state of particles $b$ and $c$. At Alice's site, she can impose four different unitary operations on particles $b$ and $c$ after she receives particle $b$ from Bob. Consequently, as for Eve, Alice's announcement on the measurement result of particles $b$ and $c$ involves sixteen possibilities of Alice and Bob's secret bits. As a result, the quantum channel includes $-\sum_{i=1}^{16} p_i \log_2 p_i = -16 \times \frac{1}{16} \log_2 \frac{1}{16} = 4$ bit information for Eve. Therefore, no information is leaked out to Eve.

## 2.3 Calculation of information-theoretical efficiency

In the above example, the $|W\rangle_{abc}$ can be used for transmitting Alice's two-bit and Bob's two-bit with two bits consumed for Alice's announcement on the measurement result of particles $b$ and $c$. Thus Cabello's information-theoretical efficiency is $\eta = \frac{4}{3+2} \times 100\% = 80\%$.

## 3 QD protocol using four-particle GHZ state
### 3.1 Protocol description

Before beginning the protocol description, let us illustrate the effect from the double CNOT operations performed on four-particle GHZ states. Four four-particle GHZ states are defined as follows:

$$|G_1\rangle_{abcd} = \frac{1}{\sqrt{2}}(|0000\rangle + |1111\rangle)_{abcd}, \tag{15}$$

$$|G_2\rangle_{abcd} = \frac{1}{\sqrt{2}}(|0001\rangle - |1110\rangle)_{abcd}, \tag{16}$$

$$|G_3\rangle_{abcd} = \frac{1}{\sqrt{2}}(|0101\rangle + |1010\rangle)_{abcd}, \tag{17}$$

$$|G_4\rangle_{abcd} = \frac{1}{\sqrt{2}}(|0100\rangle - |1011\rangle)_{abcd}. \tag{18}$$

When the double CNOT operations are performed on the above four four-particle GHZ states with particle $b$ as the control qubit and particles $c$ and $d$ as the target qubits, respectively, it follows:

$$U_{CNOT}^{bc} \otimes U_{CNOT}^{bd} \otimes |G_1\rangle_{abcd} = \frac{1}{\sqrt{2}}(|00\rangle_{ab} + |11\rangle_{ab}) \otimes |00\rangle_{cd} = |\phi^+\rangle_{ab}|00\rangle_{cd}, \tag{19}$$

$$U_{CNOT}^{bc} \otimes U_{CNOT}^{bd} \otimes |G_2\rangle_{abcd} = \frac{1}{\sqrt{2}}(|00\rangle_{ab} - |11\rangle_{ab}) \otimes |01\rangle_{cd} = |\phi^-\rangle_{ab}|01\rangle_{cd}, \tag{20}$$

$$U_{CNOT}^{bc} \otimes U_{CNOT}^{bd} \otimes |G_3\rangle_{abcd} = \frac{1}{\sqrt{2}}(|01\rangle_{ab} + |10\rangle_{ab}) \otimes |10\rangle_{cd} = |\psi^+\rangle_{ab}|10\rangle_{cd}, \tag{21}$$

$$U_{CNOT}^{bc} \otimes U_{CNOT}^{bd} \otimes |G_4\rangle_{abcd} = \frac{1}{\sqrt{2}}(|01\rangle_{ab} - |10\rangle_{ab}) \otimes |11\rangle_{cd} = |\psi^-\rangle_{ab}|11\rangle_{cd}. \tag{22}$$

Apparently, after the double CNOT operations, each of the four four-particle GHZ states collapses into two independent quantum states without any entanglement, which are composed by particles $a$ and $b$ and particles $c$ and $d$, respectively.

Suppose that both Alice and Bob have a two-bit for their own. They reach a consensus in advance that the four unitary operations $U_{00}$, $U_{01}$, $U_{10}$ and $U_{11}$ defined in section 2 also represent the classical bits 00, 01, 10 and 11, respectively.

The proposed protocol is described in detail as follows. Same to section 2, we also omit the description of security checks here. Alice produces one four-particle GHZ state randomly in one of the above four four-particle GHZ states. She sends particles $b$, $c$ and $d$ to Bob in three steps. After obtaining particles $b$, $c$ and $d$, Bob performs double CNOT operations on particles $c$ and $d$ with particle $b$ as the control qubit, respectively. Then, Bob uses $Z \otimes Z$-basis $(\{|00\rangle, |01\rangle, |10\rangle, |11\rangle\})$ to measure particles $c$ and $d$. According to Eqs.(19-22), Bob is able to know which Bell state particles $a$ and $b$ are collapsed into. On the other hand, Alice is also able to know the state of particles $a$ and $b$ after double CNOT operations, since she prepares the four-particle GHZ state by herself. Then, Bob encodes his own two-bit on particle $b$ by performing one of the four unitary operations $\{U_{00}, U_{01}, U_{10}, U_{11}\}$. Afterward, Bob sends particle $b$ to Alice. Alice encodes her own two-bit on particle $a$ by performing one of the four unitary operations $\{U_{00}, U_{01}, U_{10}, U_{11}\}$. Then, Alice uses Bell-basis to measure particles $a$ and $b$, and announces her measurement result to Bob. According to his own unitary operation, Bob can extract Alice's two-bit from her announcement on the measurement result of particles $a$ and $b$. Analogously, Alice can also obtain Bob's two-bit through her own unitary operation.

A concrete example is given to further explain the above protocol. Assume that the four-particle GHZ state Alice prepares is $|G_3\rangle_{abcd}$. Moreover, suppose that Alice and Bob want to transmit bits **01** and **00** to each other, respectively. After obtaining particles $b$, $c$ and $d$, Bob performs double CNOT operations on particles $c$ and $d$ with particle $b$ as the control qubit, respectively. Consequently, the $|G_3\rangle_{abcd}$ evolves in a manner shown as Eq.(21). Then Bob uses $Z \otimes Z$-basis to measure particles $c$ and $d$ to know the state of particles $a$ and $b$. Afterward, Bob imposes $U_{00}$ on particle $b$ and sends it to Alice. Then, Alice performs $U_{01}$ on particle $a$. Naturally, particles $a$ and $b$ evolve as follows:

$$|\psi^+\rangle_{ab} \Rightarrow U_{01}^a \otimes U_{00}^b \otimes |\psi^+\rangle_{ab} = |\phi^+\rangle_{ab}. \tag{23}$$

Afterward, Alice performs Bell-basis measurement on particles $a$ and $b$, and announces her measurement result to Bob. Bob can read out that Alice's two-bit is **01** from her announcement. Analogously, Alice can also infer out that Bob's two-bit is **00**.

### 3.2 Security analysis

The information leakage problem of the above protocol is analyzed here. After receiving particles $b$, $c$ and $d$ from Alice, Bob performs double CNOT operations on particles $c$ and $d$ with particle $b$ as the control qubit, respectively. Bob is able to know which Bell state particles $a$ and $b$ are collapsed into. On the other hand, since Alice prepares the $|G_3\rangle_{abcd}$ by herself, she can automatically know the state of particles $a$ and $b$ after double CNOT operations from Eq.(21). In this way, the state of particles $a$ and $b$ after double CNOT operations is shared privately between Alice and Bob. Naturally, Eve has no access to the state of particles $a$ and $b$ after double CNOT operations. Therefore, as for Eve, Alice's announcement on the measurement result of particles $a$ and $b$ involves sixteen possibilities of Alice and Bob's secret bits. As a result, the quantum channel includes $-\sum_{i=1}^{16} p_i \log_2 p_i = -16 \times \frac{1}{16} \log_2 \frac{1}{16} = 4$ bit information for Eve. Therefore, no information is leaked out to Eve.

### 3.3 Calculation of information-theoretical efficiency

In the above example, the $|G_3\rangle_{abcd}$ can be used for transmitting Alice's two-bit and Bob's two-bit with two bits consumed for Alice's announcement on the measurement result of particles $a$ and $b$. Thus Cabello's information-theoretical efficiency is

$$\eta = \frac{4}{4+2} \times 100\% = 66.7\%.$$

## 4 Comparisons of previous information leakage resistant QD protocols

We compare our protocols with the previous information leakage resistant QD protocols[17-23] on three aspects including the initial quantum resource, the quantum measurement and Cabello's information-theoretical efficiency. Table 1 summarizes the comparison results. It can be concluded from Table 1 that, compared with the protocols in Refs.[17-22], the advantage of our protocols with Bell state and W state lies in Cabello's information-theoretical efficiency.

Table 1. Comparisons of previous information leakage resistant QD protocols

| | The protocol of Ref.[17] | The protocol of Ref.[18] | The protocol of Ref.[19] | The protocol of Ref.[20] | The protocol of Ref.[21] | The protocol of Ref.[22] | The protocol of Ref.[23] | The proposed protocol with Bell state | The proposed protocol with W state | The proposed protocol with four-particle GHZ state |
|---|---|---|---|---|---|---|---|---|---|---|
| Initial quantum resource | Bell states | single particles | Bell states and single particles | Bell states | GHZ states | Bell states | Nearly single particles | Bell states | W states | four-particle GHZ states |
| Quantum measurement | Bell-basis measurements | single-particle measurements | single-particle measurements and Bell-basis measurements | Bell-basis measurements | GHZ-basis measurements | Bell-basis measurements | single-particle measurements | single-particle measurements | single-particle measurements and Bell-basis measurements | single-particle measurements and Bell-basis measurements |
| Cabello's information-theoretical efficiency | 66.7% | 66.7% | 75% | 66.7% | 66.7% | 66.7% | Nearly 100% | 100% | 80% | 66.7% |

## 5 Conclusion

To summarize, we propose a kind of QD protocols without information leakage, which need the assistance of auxiliary quantum operation. The participant in charge of preparation can automatically know the collapsed states after quantum operation performed on the prepared quantum states. The other participant is able to know the collapsed states derived from the prepared quantum states through quantum measurement. The information leakage problem is avoided by means of imposing auxiliary quantum operation on the prepared quantum states.

**Acknowledgements**

Funding by the National Natural Science Foundation of China (Grant Nos.61402407, 11375152) is gratefully acknowledged.